\documentclass[runningheads]{llncs}
\usepackage{graphicx}
\usepackage{adjustbox}
\usepackage{breakcites}
\usepackage{url}

\usepackage[breaklinks]{hyperref}
\usepackage{lipsum}

\begin{document}

This is a pre-print version of the chapter published in Advances in Bias and Fairness in Information Retrieval (BIAS 2021). The  final  authenticated  version  is available online at \href{https://doi.org/10.1007/978-3-030-78818-6_5}{https://doi.org/10.1007/978-3-030-78818-6\_5}.



To cite this chapter: 

Makhortykh M., Urman A., and  Ulloa R. (2021) Detecting Race and Gender Bias in Visual Representation of AI on Web Search Engines. In L. Boratto, S. Faralli, M. Marras, and G. Stilo (Eds.), Advances in Bias and Fairness in Information Retrieval (pp. 36-50). Springer. DOI: \href{https://doi.org/10.1007/978-3-030-78818-6_5}{https://doi.org/10.1007/978-3-030-78818-6\_5}.

\newpage


\title{Detecting race and gender bias in visual representation of AI on web search engines}

\titlerunning{Visual representation of AI in web search}

\author{Mykola Makhortykh\inst{1} \and
Aleksandra Urman\inst{1,2} \and
Roberto Ulloa\inst{3}}

\authorrunning{Makhortykh et al.}

\institute{University of Bern, Switzerland \and University of Zurich, Switzerland \and
GESIS - Leibniz Institute for the Social Sciences, Cologne, Germany}

\maketitle              

\begin{abstract}
Web search engines influence perception of social reality by filtering and ranking information. However, their outputs are often subjected to bias that can lead to skewed representation of subjects such as professional occupations or gender. In our paper, we use a mixed-method approach to investigate presence of race and gender bias in representation of artificial intelligence (AI) in image search results coming from six different search engines. Our findings show that search engines prioritize anthropomorphic images of AI that portray it as white, whereas non-white images of AI are present only in non-Western search engines. By contrast, gender representation of AI is more diverse and less skewed towards a specific gender that can be attributed to higher awareness about gender bias in search outputs. Our observations indicate both the the need and the possibility for addressing bias in representation of societally relevant subjects, such as technological innovation, and emphasize the importance of designing new approaches for detecting bias in information retrieval systems.

\end{abstract}

\section{Introduction}
Web search engines are important information intermediaries that help users navigate through web content. By filtering and ranking information in response to user queries, search engines determine what users learn about specific topics or entities \cite{kulshrestha_search_2019} in turn influencing individual and collective perception of social reality \cite{gillespie_relevance_2014}. However, search engine outputs can be biased - that is systematically skewed towards particular individuals or groups \cite{friedman_bias_1996} - which may lead to the distorted perception of the search subject and potentially result in negative societal effects such as racial or gender discrimination.

A growing number of studies discusses how search engines perpetuate biases related to gender and race, in particular in image search results \cite{kay_unequal_2015,noble_algorithms_2018,otterbacher_competent_2017}. Because of their affective and interpretative potential \cite{bleiker_visual_2018}, images can be effective means of educating the public about complex social phenomena, such as gender or race, but also of reiterating stereotypes \cite{kay_unequal_2015}. With image search being used for a broad range of purposes, varying from educators preparing teaching materials \cite{muller_health_nodate} to media professionals producing new content \cite{otterbacher_investigating_2018}, its biased outputs can reinforce skewed representation, in particular of already vulnerable groups, and amplify discrimination \cite{noble_algorithms_2018}.

Currently, research on race and gender bias in image search focuses on visual representation of a few subjects, such as professional occupations \cite{kay_unequal_2015} or emotions \cite{otterbacher_investigating_2018}. However, there is a growing recognition that representation of other aspects of contemporary societies can also be genderly or racially skewed. One of such aspects is technological innovation, the representation of which in the West historically tended to be decontextualized and often associated with masculinity \cite{blake_rethinking_2005} and whiteness \cite{katz_artificial_2020}. Such biases can further aggravate existing inequalities by influencing hiring decisions (e.g., by stereotyping a certain field as racially homogeneous) and positioning the technologies, predominantly portrayed as White, above marginalised non-white people \cite{cave_whiteness_2020}. Biases found to be present in web search outputs (e.g., \cite{noble_algorithms_2018,kay_unequal_2015}) have the potential to influence public opinion and perceptions of the social reality \cite{epstein_search_2015,kulshrestha_search_2019}. This is further aggravated by the fact that users tend to trust the output of search engines \cite{schultheis_we_2018}

Besides expanding the current focus of search bias research to new areas, it is also important to consider the consequences of recent studies on search engine auditing for evaluating the robustness of bias measurements. Firstly, while the effect of personalization on the variability of search outputs is shown to be minor \cite{unkel_googling_2019,trielli_partisan_2020}, the influence of randomization (e.g., result reshuffling for maximizing user engagement) can be more significant \cite{makhortykh_how_2020} and is yet to be accounted for in the context of bias research. Second, despite substantial differences in content selection across search engines \cite{jiang_business_2014,a2021mitigating,makhortykh_how_2020}, the majority of existing research focuses on individual search engines (e.g., Google\cite{kay_unequal_2015} or Bing\cite{otterbacher_investigating_2018}), whereas possible bias variation between different engines (including the ones prevalent in non-Western context, such as Yandex or Baidu) remains understudied. 

In this paper, we aim to make two contributions. First, we introduce a mixed-method approach for detecting race and gender bias in image search outputs that takes into consideration potential effects of randomization and personalization. Second, we apply this method for conducting a cross-engine comparison of bias in the visual representation of artificial intelligence (AI). Our choice of a case study is attributed to common criticism of AI representation being racially and genderly skewed both in popular culture and industry \cite{sparrow_robots_2020, adams_helen_2020} and the recent claims about Google amplifying these biases via its image search results \cite{sparrow_robots_2020,cave_whiteness_2020}.

\section{Related work: race and gender bias in image search}
The possibility that outputs of web search engines can be systematically skewed towards certain gender and racial groups is increasingly recognized both by the broad public and information retrieval (IR) community \cite{otterbacher_investigating_2018}. Race and gender biases are found in different IR systems associated with search engines, including text search results \cite{pradel_biased_2020,noble_algorithms_2018} and search autocompletion \cite{baker_why_2013,bonart_investigation_2019}. However, image search is particularly relevant in this context, because of high interpretative and affective potential of visual information \cite{bleiker_visual_2018, makhortykh2020memory} that makes it a potent means of challenging, but also forming stereotypes.

Despite the growing recognition of the problem, there are still relatively few studies which look at biases in image search outputs in a systematic way. To study the prevalence of gender bias, Kay et al. \cite{kay_unequal_2015} collected the US Bureau of Labor and Statistics data on gender distribution per occupation and compared it with results of Google image search for the respective occupations. Their findings indicate that Google image search outputs tend to exaggerate gender biases in relation to occupations viewed as male- or female-dominated. Otterbacher et al. \cite{otterbacher_competent_2017} used Bing image search API to extract results for "person" query and then classified them using Clarifai. Their findings indicate that male images occur more commonly than female ones for gender-neutral queries; furthermore, search results tend to present males as more competent and purpose-oriented.

Even less research was done on racial bias in image search outputs. Using a selection of racialized and gendered web search queries (e.g., “black girls”), Noble \cite{noble_algorithms_2018} employed qualitative content analysis to identify multiple cases when Google promoted racist and misogynistic representation of women and minority groups. Araujo et al. \cite{araujo_identifying_2016} used feature extraction to compare outputs from Google and Bing image search APIs for “ugly woman” and “beautiful woman” queries and found that images of black women were more often identified as “ugly”, whereas white women were positively stereotyped.

\section{Case study: Racial and gender representation of AI}
The ongoing recognition of the complex relationship between technical (e.g., algorithms) and social (e.g., race) constructs has substantial implications for how modern technology is perceived. In the case of AI, which can be broadly defined as the ability of human-made artifacts to engage in intellectual behavior \cite{nilsson_artificial_1998}, this connection is particularly strong. Its strength is attributed both to conceptual reasons, such the bilateral relationship between intellectual behavior and social norms/activities \cite{doise2013social}, and the increasing adoption of AI for the tasks dealing with societal matters (e.g., predictive policing \cite{eubanks2018automating}). 

The tendency to anthropomorphize AI - that is to present it in a human-like form either physically or digitally \cite{cave_whiteness_2020,leufer_why_2020} - further problematizes its relationship with social constructs. There are multiple theories explaining the historical tendency to integrate anthropomorphic features in product design \cite{disalvo_seduction_2003}, but generally anthropomorphism is an important factor in making complex technology more familiar and comfortable to use. However, anthropomorphism also stimulates application of social categories (e.g., race or gender) to the non-human entities, such as AI, and it has substantial implications both to their perception and representation \cite{bartneck_robots_2018}. 

The racial representation of AI in the Western Anglophone culture is characterized by whiteness both in terms of physical appearance and behavior. Cave and Dihal \cite{cave_whiteness_2020} list multiple cultural products in which AI is presented as exclusively white (Terminator, Blade Runner, I, Robot to name a few). While the portrayal of AI in recent cultural products slowly becomes more diverse (e.g., Westworld and Humans), it is still predominantly treated as white. 

Similar to popular culture, the historical and institutional context of AI industry in the West is argued to be related to whiteness \cite{katz_artificial_2020}. Besides multiple other consequences, it affects how AI is imagined and represented as indicated by the prevalent use of white materials and surfaces for constructing robots \cite{sparrow_robots_2020} and the reliance on sociolinguistic markers associated with whiteness (e.g., by omitting dialects related to non-white groups when developing conversational agents \cite{cave_whiteness_2020}.     

In contrast to the limited variety of its racial representation, the gender representation of AI is more diverse (albeit still quite binary). In the case of popular culture, there are multiple instances of portraying AI as male and female entities. The number of fictitious representations of AI as female (Metropolis, Her, Ghost in the Shell, Ex Machina) might be even higher than the number of male ones (A.I., I, Robot, Prometheus). At the same time, many of these representations portray female AI just as servants to their (usually male) masters, who often treat AI as a means of satisfying their (sexual) needs \cite{adams_helen_2020}. 

A similar relationship between AI and gender can be observed in the industry, where the most well-known AI-based assistants (e.g., Cortana or Siri) have female features. While the industrial treatment of gender aspects of AI does not necessarily rely on its intense sexualization as much as popular culture, it still iterates the notion of women being subordinate to men and intended to be used by their masters \cite{adams2019addressing}. 

\section{Methodology}
To collect data, we utilized a set of virtual agents - that is software simulating user browsing behavior (e.g., scrolling web pages and entering queries) and recording its outputs. The benefits of this approach, which extends algorithmic auditing methodology introduced by Haim et al. \cite{haim_abyss_2017}, is that it allows controlling for personalization \cite{hannak_measuring_2013} and randomization \cite{makhortykh_how_2020} factors influencing outputs of web search. In contrast to human actors, virtual agents can be easily synchronized (i.e., to isolate the effect of time at which the search actions are conducted) and deployed in a controlled environment (e.g., a network of virtual machines using the same IP range, the same type of operating system (OS) and the same browsing software) to limit the effects of personalization that might lead to skewed outputs. 

In addition to controlling for personalization, agent-based auditing allows addressing randomization of web search that is caused by search engines testing different ways of ranking results to identify their optimal ordering for a query (e.g., the so-called “Google Dance” \cite{battelle_search_2011}). Such randomization leads to a situation, when identical queries entered under the same conditions can result in different sets of outputs (or their different ranking), thus making the observations non-robust. One way of addressing this issue is to deploy multiple virtual agents that simultaneously enter the same search query to determine randomization-caused variation in the sets of outputs that can then be merged into a single, more complete set. 

For the current study, we built a network of 100 CentOS virtual machines based in the Frankfurt region of Amazon Elastic Compute Cloud (EC2). On each machine, we deployed 2 virtual agents (one in Chrome browser and one in Mozilla Firefox browser), thus providing us with 200 agents overall. Each agent was made of two browser extensions: a tracker and a bot. The tracker collected the HTML and the metadata of all pages visited in the browser and immediately sent it to a storage server. The bot emulated a sequence of browsing actions that consisted of (1) visiting an image search engine page, (2) entering the “artificial intelligence” query, and (3) scrolling down the search result page to load at least 50 images. 

Before starting the emulation, the browsers were cleaned to prevent the search history affecting the search outputs. While there is a possibility that search engines infer the gender of the agent based on the previous IP behavior (i.e., the use of the IP by other AWS users before it was assigned to a virtual agent deployed as part of our experiment), we expect that the use of multiple agents shall counter this potential limitation, because it is unlikely that the large number of randomly assigned IPs will be associated with one specific gender. 

The study was conducted on February 27, 2020. We distributed 200 agents between the world’s six most popular search engines by market share: Google, Bing, Yahoo, Baidu, Yandex, and DuckDuckGo (DDG) \cite{statcounter_search_2020}. For all engines, the ".com" version of the image search engine was used (e.g., google.com). The agents were equally distributed between the engines; however, because of technical issues (e.g., bot detection mechanisms), some agents did not manage to complete their routine. The overall number of agents per engine which completed the full simulation routine and returned the search results was the following: Baidu (29), Bing (30), DDG (34), Google (33), Yahoo (31), and Yandex (21).

For our analysis, we extracted from collected HTML links related to top 30 image search results for each agent. Then, we divided these links into three subgroups: 1) results from 1 to 10; 2) results from 11 to 20; and 3) results from 21 to 30. Such a division allowed us to investigate differences in terms of race and gender bias between top results (i.e., 1-10), which are usually the only ones viewed by the users \cite{pan_google_2007}, and later results. Then, we aggregated all images for each subgroup per engine to account for possible randomization of search outputs on individual agent level, and removed duplicate images. The number of unique images in each subgroup per each engine is shown in Table \ref{tab:table1}; the numbers do not include images which were not accessible anymore (e.g., because of being removed from the original websites).

\begin{table}[ht]
\caption{The number of unique images per each result subgroup per engine
}~\label{tab:table1}
\centering
\begin{adjustbox}{width=0.8\textwidth}
\small
\begin{tabular}{|l|l|l|l|l|l|l|}
\hline
              & Baidu & Bing & DuckDuckGo & Google & Yahoo & Yandex \\ \hline
Results 1:10  & 10    & 16   & 11  & 11     & 13    & 12     \\ \hline
Results 11:20 & 12    & 10   & 11  & 16     & 15    & 14     \\ \hline
Results 21:30 & 12    & 10   & 10  & 13     & 16    & 15     \\ \hline
\end{tabular}
\end{adjustbox}
\end{table}

To detect race and gender bias in search outputs, we relied on their manual coding. While some earlier studies (e.g., \cite{otterbacher_competent_2017}) use image recognition for extracting image features, its applicability for bias detection has been questioned recently \cite{schwemmer_diagnosing_2020} considering the possibility of recognition approaches being biased themselves. Hence, we used two coders to classify all the collected images based on categories listed below.  To measure intercorder reliability, we calculated Kripperndorf’s alpha values that showed an acceptable level of reliability: 0.73 (antropomorphism), 0.69 (race), 0.67 (sex). Following the reliability assessment, the identified disagreements were resolved by the original coders using consensus-coding.

\textit{Anthropomorphism:} We determined whether the image of AI includes any anthropomorphic elements, such as human(-like) figures or parts of human bodies. Depending on their exact appearance, such elements can indicate what human-like features are attributed to AI and how its developers and users are portrayed. Most importantly, this category determines the subset of images which can be subjected to gender and race bias, because both forms of bias are primarily applicable to anthropomorphic portrayals of AI. 

\textit{Race:} For anthropomorphized images, we identified the race of the portrayed entity to determine whether there is racial skew in AI representation. Following Cave and Dihal \cite{cave_whiteness_2020}, we treat racialized representation in broad terms and interpret the coloring of AI elements as a form of racial attribution. Hence, both a white-skinned individual and a human-like android made of white material can be treated as White. The options included 1) white, 2) non-white, 3) mixed  (when both white and non-white entities were present), 4) abstract (when an entity can not be attributed to any human race), and 5) unknown (when it was not possible to reliably detect race). 

We acknowledge that treating race as a binary (white/non-white) category is a simplification as the complex notion that ignores multiple nuances, in particular different categories of non-white population (e.g., Black, Hispanic or Asian). However, numerous recent explorations of the role of race in the societal hierarchies as well as media-driven racial bias use the same white/non-white dichotomy \cite{eddo-lodge_why_2020,hubinette_be_2009,kivel_uprooting_2017,heider_white_2014}, in part to avoid shifting the focus from the uniqueness of white privilege compared with various degrees of marginalization of non-white population. The same logic can be applied to the study of technological innovation, in particular in the context of AI, which has been historically related to whiteness \cite{katz_artificial_2020}. Furthermore, it is difficult to identify different sub-categories of non-white population using only visual cues, considering the complex notion of race and the fact that is not necessarily based on one's appearance only. Because of these reasons, we believe that in the context of the present study, the binary categorization of race is suitable despite its limitations. 

\textit{Sex:} For anthropomorphized images, we determined the sex of the entity portrayed to determine whether there is a gendered skew. We used sex as a proxy for gendered representation because of the complexity of the notion of gender. Unlike sex, which is a binary concept, gender encompasses a broad variety of social and cultural  identities that makes it hard to detect based on visual cues. Hence, we opted out for a more robust option that is still sufficient for evaluating gender-related aspects of AI representation. The possible options included 1) male, 2) female, 3) mixed (when both male and female entities were present), 4) abstract (when an entity was shown as sexless), and 5) unknown (when it was not possible to reliably detect sex).

\section{Findings}
\subsection{AI and antropomorphism}
Unlike Kay et al. \cite{kay_unequal_2015}, who had data on gender distribution for occupations to compare their representation in image search outputs, we do not have a clear baseline for AI representation. Hence, we follow Otterbacher et al. \cite{otterbacher_competent_2017} and treat the unequal retrievability - that is the accessibility of outputs with specific characteristics \cite{traub_querylog-based_2016} - as an indicator of bias in search outputs. By systematically prioritizing images with specific features  (e.g., the ones showing males and not females; \cite{otterbacher_competent_2017}), the system creates a skewed perception of the phenomenon represented via its outputs. 

\begin{figure}[h]
  \centering
  \includegraphics[width=0.8\linewidth]{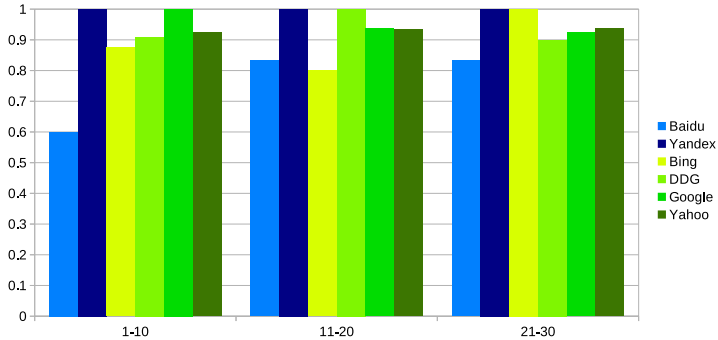}
  \caption{Ratio of anthropomorphic representations of AI per result set for each engine (1-10 refers to results 1 to 10; 11-20 to results 11 to 20; 21-30 to results 21 to 30).}
  \label{fig1}
\end{figure}

The first sign of the skewed representation of AI by search engines is the tendency for its anthropomorphization (Fig.\ref{fig1}). The proportion of images showing AI not as an anthropomorphized entity constitutes less than 10\% for most engines with a single exception coming from Baidu, the largest Chinese search engine. On Google and Yandex, all images appearing in the top 10 results show AI as a human-like entity.  For other Western engines, the proportion of anthropomorphized AI images increased after the first 10 results, and in some cases (Bing, DDG) also reached 100\% for the second and the third sets of outputs.

The anthropomorphized representations of AI usually take one of two forms: a schematic representation of a human brain or a human-like figure made of different materials (e.g., metal, plastic, or pixels). The way of presenting AI as “shiny humanoid robots” \cite{leufer_why_2020} can be attributed to it being the most recognizable way of presenting it in Western popular culture. However, by reiterating human-like AI images, search engines also create more possibilities for bias compared with more schematic or non-anthropomorphized representations.

\subsection{AI and race}
\begin{figure}[h]
  \centering
  \includegraphics[width=1.0\linewidth]{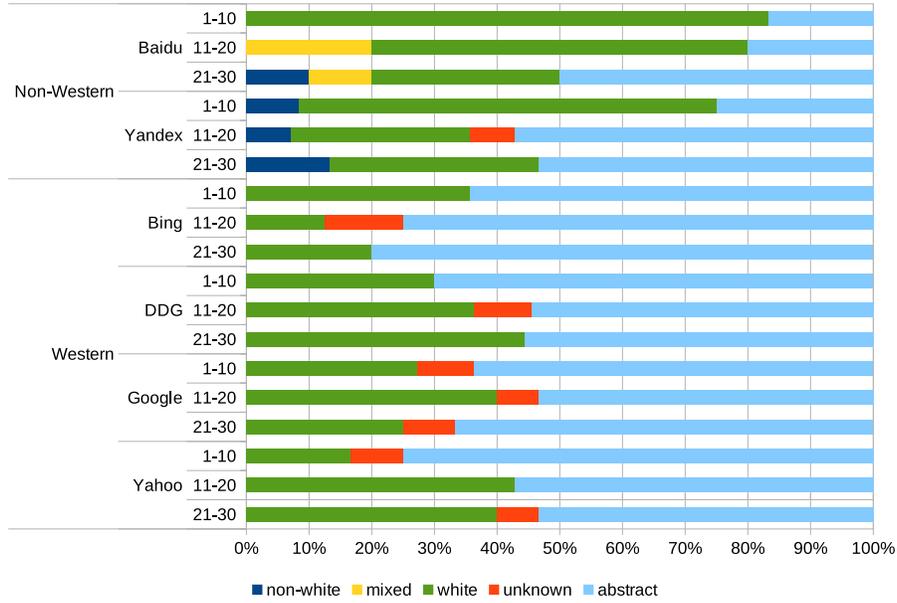}
  \caption{Proportion of racialized representations of AI among anthropomorphized images  (1-10 refers to results 1 to 10; 11-20 to results 11 to 20; 21-30 to results 21 to 30).}
  \label{fig2}
\end{figure}

Our analysis showed that non-racialized portrayals of AI are prevalent on Western search engines (Fig.\ref{fig2}). With the exception of Bing, where racialized images prevail among the first 10 results, Western engines tend to put more abstract images (e.g., schematic brain images or bluish human figures) on the top of search results, whereas later sets of outputs become more skewed towards specific racial representations. However, a different pattern is observed on non-Western engines, where racialized images of AI appear on the top of search results and become less visible in the later outputs.

In those cases, when AI images have a racial component, it is almost always associated with whiteness, thus supporting earlier claims about a skewed racial portrayal of AI on search engines \cite{cave_whiteness_2020,sparrow_robots_2020}. The most common form such an association takes is the stylization of human-like entities representing AI with white materials and Caucasian face features. In some cases, it is supplemented with images of humans or part of their bodies (e.g., arms) representing  developers or users of AI, most of whom are again shown as white. While the proportion of such racialized images in top 10 results for Western search engines is relatively low, their presence still promotes “White utopian imaginary” \cite{cave_whiteness_2020} of AI and enforces racial homogeneity in relation to it. 

Such a skewed representation is amplified by an almost complete absence of non-white entities in  image search results. Their absence is particularly striking for Western search engines, where the only form of racial diversity available are images of body parts that can not be easily related to a particular race (e.g., an arm the color of which can not be easily detected because of the way the image is lit). This exclusive focus on whiteness can be treated as a form of symbolic erasure of non-white groups not only from AI industry, but also the larger context of its use that encompasses most of contemporary societies \cite{cave_whiteness_2020}.

Surprisingly, the non-Western engines, which prioritize images accentuating AI whiteness more than the Western ones, are also the only ones to include non-white or mixed AI portrayals. With the exception of two images showing AI as a black- or brown-skinned entity, these images show non-white AI developers or users. Such discrepancy leads to a situation, where the use of AI is contextualized by adding non-white contexts, but the core interpretation of AI as a "technology of whiteness" \cite{katz_artificial_2020} is not challenged.  

\subsection{AI and gender}
Similarly to racial representation, we observed the prevalence of non-gendered portrayals of AI. The tendency for not attributing specific sex features to it was even more pronounced than in the case of race, where the proportion of abstract representations was lower. Just as in the case of race, the majority of Western search engines (except Bing) prioritized gender-neutral images among the top 10 results with more gendered images appearing for lower results. The opposite pattern was again observed for Yandex (but not Baidu) and Bing, where the top 10 results contained more gendered images than the later ones.

In the case of gendered images of AI, we did not observe that much of a discrepancy between different groups as in the case of racialized ones. With the exception of Bing, Western search engines included in the top 10 results images based on which it was not possible to clearly identify the entity’s sex (e.g., human arms which could belong both to males and females). Later sets of results also included images presenting AI as female entities, but their proportion rarely exceeded 10\% of search outputs.

\begin{figure}[h]
  \centering
  \includegraphics[width=1.0\linewidth]{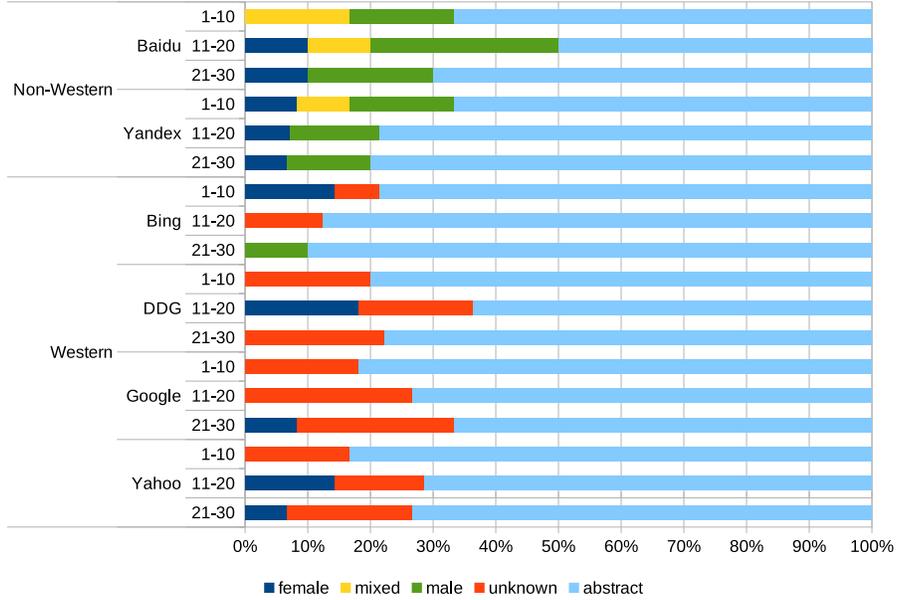}
  \caption{Proportion of gendered representations of AI among anthropomorphized images (1-10 refers to results 1 to 10; 11-20 to results 11 to 20; 21-30 to results 21 to 30).}
  \label{fig3}
\end{figure}

As in the case of racialized images, non-Western engines provided more diverse representations of AI, including both female and male entities as well as the mix of two. Unlike Western engines, where images of AI as a male entity were almost absent, both Baidu and Yandex prioritized more masculine representation of AI. Such effect was achieved by highlighting images of both male developers and users as well as human-like robots with masculine facial features. One possible explanation of non-Western engines promoting a masculine portrayal of AI can be its different representation in popular culture. At least in the case of Russia, where Yandex originates from, a number of prominent cultural products present AI as a masculine entity (e.g., The Adventures of Electronic, Far Rainbow, Guest from the Future), whereas feminine representations are rather few.

While the aforementioned observation can be treated as evidence that search engine outputs depend on popular culture representations of AI, we did not observe any overly sexualized images of female AI despite its intense sexualization in Western popular culture. This finding indicates that cultural embeddedness of bias does not necessarily translate into its visibility in search outputs and can be attributed to more active countering of gender bias in the recent years \cite{noble_algorithms_2018}.

\section{Discussion}
Our observations indicate that visual representation of AI on the world's most popular search engines is skewed in some racial and, to a lesser degree, gender aspects. While it is not sufficient to claim that search mechanisms used to retrieve information about AI are racially or genderly biased, our findings support earlier research \cite{otterbacher_competent_2017} that found search engines reiterating social biases. In the case of AI, it results in predominantly white portrayal of the technology and the omittance of non-white AI designs as well as non-white developers and users. By offering rather skewed selection of visual information, search engines misrepresent important developments in the field of AI and erase the presence of non-white groups that can be viewed as a form of discrimination.

Similar to other forms of web bias \cite{baeza2018bias}, the white-centric representation of AI on search engines can be explained by multiple factors. Because of its prevalence in Western Anglophone popular culture and industry, representation of AI as White commonly appears on "authoritative" websites, such as the ones related to government and research institutions and mainstream media. Outputs from these websites are prioritized both because they are treated as more reliable sources of information \cite{grind2019google} and because they often have the large number of backlinks, a feature which is important for website ranking on the majority of search engines \cite{noauthor_search_nodate} (Yandex, however, is a notable exception with its larger emphasis not on backlinking, but on user engagement \cite{noauthor_9_nodate}).

Additional factor which contributes to racial bias in AI representation is the fact that image search outputs are often based on text accompanying the image, but not on the image features \cite{cui2008real, noauthor_yandex_nodate}. Under these circumstances, the search algorithm is not necessarily able to differentiate between white and non-white representations of AI. Instead, it just retrieves images which are accompanied by certain text from the websites, the ranking of which is determined using the same criteria as text search results. Considering that racial bias in AI representation remains mainstream \cite{cave_whiteness_2020} as contrasted by gender bias (e.g., it is harder to imagine academic or government websites hosting images of sexualized AI), it results in the iteration of white-centric representations, in particular by Western search engines.  

The reliance on textual cues for generating image search outputs and engine-specific ranking signals (e.g., number of backlinks and source type) can also explain differences in AI representation between Western and non-Western search engines. Unlike Western engines, where the selection of ranking signals is similar and results in reiteration of the same set of images stressing the whiteness of AI, the focus on the specific regions (i.e., China for Baidu and Russia for Yandex) together with substantial differences in ranking mechanisms (e.g., prioritization of backlinks coming from China for Baidu \cite{noauthor_search_nodate} and the reliance on different ranking signals for Yandex \cite{noauthor_9_nodate}) leads to the inclusion of more non-white representations of technology. However, if this explanation is valid, then in order to be able to deal with racial bias in a consistent manner, search engines would need either to more actively engage with actual image features (and not just text accompanying images) or expand the selection of websites prioritized for retrieving image outputs beyond currently prioritized mainstream Western websites, where white-centered AI representations are prevalent.

Overall, racial bias in the way web search mechanisms treat visual representation of AI can hardly be viewed as something that search engines invent on their own. However, they do reinforce the bias by creating a vicious cycle in which images of AI as "technology of whiteness" \cite{katz_artificial_2020} appear on the top of search results and are more likely to be utilized by users, including educators or media practitioners. However, this reinforcement loop can be broken as shown by the substantially less biased representation of AI in terms of gender: despite the strong tendency for its femalization and subsequent sexualization in popular culture, we found relatively few gendered images of AI in the top results and none of them was sexualized. 

Together with the earlier cases of addressing skewed web search outputs that were identified by the researchers (e.g., racialized gender bias \cite{noble_algorithms_2018}), our observations support the argument of Otterbacher \cite{otterbacher2018addressing} about the importance of designing new approaches for detecting bias in IR systems. In order to be addressed, bias has first to be reliably identified, but so far there is only a few IR studies that investigate the problem in a systematic way. By applying a new approach to examine bias in the context of AI, our paper highlights the importance of conducting further research to achieve better understanding of how significant are racial and gender biases in search outputs in relation to different aspects of contemporary societies, including (but not limited to) other forms of innovation.

It is also important to note several limitations of the research we conducted. First, we used very simple binary classification schematas both for race and gender features of AI portrayal. Second, our observations rely on a snapshot experiment conducted at a certain point of time, so it does not account for possible fluidity of image search results. Third, the experimental setup (i.e., the choice of the query and the infrastructure location) can also influence the observations produced.   

\bibliographystyle{splncs04}
\bibliography{BIAS2021}

\end{document}